\documentclass[12pt]{article}
\usepackage{geometry}
\usepackage[utf8]{inputenc}
\usepackage{amsmath}
\usepackage{graphicx}
\usepackage{color}
\usepackage{amsfonts}

\usepackage{amstext}
\usepackage{amssymb}
\usepackage{babel}
\usepackage[colorlinks,allcolors=red]{hyperref}
\begin{document}
\title{Does  Quantum Mechanics apply to Macroscopic Objects? How to define  "Macro"?}

\author{Ladislaus Alexander B\'anyai}

\maketitle
{\sl Institut f\"ur Theoretische Physik, Goethe-Universit\"at, Frankfurt am Main  
}
\\

{\bf banyai@itp.uni-frankfurt.de} {\color{blue}corresponding author} 

ORCID ({\color{blue}https://www.orcid.org/0000-0002-0698-4202)}

\abstract{ According to Quantum Mechanics a narrow wave-packet of the center of mass of any macroscopic object should spread out after some time. The problem is usually waved out by assuming that due to their heavy masses this occurs  over astronomical times. Without a clear definition of macroscopic objects this remains ambiguous. On the other hand,
Quantum Mechanics allows and energetically even prefers largely spread out c.m. states that have never
been seen. Why is this the real state of the world we know? Does it suggest a micro-macro threshold
with different theoretical descriptions? Does  micro-miniaturization offer  hope for answering these questions?   }

\vspace{1cm}

\section{Introduction}

One of the strange features of Quantum Mechanics is that for its formulation, one needs the classical physics the same one expects to emerge as its macroscopic limit. All experiences with quantum objects have to be analysed through classical "glasses". On the other hand, classical theory describes well the behaviour of macroscopic objects around us. But what is the actual definition of a macroscopic object? Where is the threshold (if any) between the microscopic and the macroscopic worlds?

One should bear in mind that Quantum Mechanics is just a theoretical frame
to describe the properties of a system that may be treated as being isolated and having a finite number of degrees of freedom. (Strictly speaking, there are no isolated physical systems and the number of degrees of freedom of a system depends on our modelling it.) The basic elements are the concept of the wave function (quantum state), the Schr\"odinger equation for its evolution, the observables (self adjoint operators) and the probabilistic interpretation of the measurement. The system itself must be defined first from the experimental point of view and this is based on our general vague conception and long experience about the surrounding world. Then one associates a certain Hamiltonian to it and verifiable mathematical predictions of Quantum Mechanics emerge.

It is a general accepted knowledge  that the hydrogen atom consist of a positively charged proton and a negatively charged electron. Its classical description could not explain neither the stability of the atom nor the discrete spectrum of emitted-absorbed light. Quantum Mechanics uses this model as a starting point for the definition of the Hamilton operator in the Schrödinger equation to get concordance with experiment, but this is far away from the picture surviving in public's mind. 

 Quantum Mechanics on its own says nothing about its object or the structure of matter. It is an exceptionally successful theory, nevertheless the relationship between the macroscopic classical and the microscopic quantum mechanical description of matter is still subject of worry.

In the first half of the twentieth century the big successes of quantum theory were obtained mainly in the field of ("elementary") microscopic particles: electrons, atoms, nuclei. Thereafter, the most spectacular evolution occurred in understanding the properties of condensed matter based on
 many-body quantum mechanics. This step revolutionized our technology.

 Now we witness another
exceptional experimental development in producing ever smaller pieces of solid matter (semiconductor chips, micro-structures). We are approaching the
microscopic world from above. Therefore, the answer to the question about the  upper end of the "quantum-domain", might be very close.

I want here  to underline some simple but often ignored aspects in this context. The theory of condensed
matter is conceived as the quantum mechanical description of bound states of an enormous number
of Coulomb interacting electrons and ions. This picture is very successful. In the absence of an external potential the center of mass
(c.m.) motion is fully separable from the relative (internal) one \cite{Messiah} and its wave function satisfies the
free Schr\"odinger equation of a particle with the total mass  $M$ of the system. One applies Quantum Mechanics
however, only to the relative motion, while the center of mass motion is considered irrelevant and left to
the classical description. 

In what follows,  we shall discuss only the center of mass motion.
In the real world the center of mass states of the macroscopic objects surrounding us are from the quantum mechanical point of view sharply defined wave
packets. According to the Schr\"odinger equation these should spread out in time.
A somewhat appeasing argument is given in the frame of Ehrenfest’s theorem in almost all handbooks
about their stability for macroscopic (heavy) masses. However, no satisfactory  definition  of the term "macroscopic" was given until now.

In Section 2 we recapitulate the standard quantum mechanical description of an ideal wave
packet (here the free motion of the c.m) and pose some basic questions. in Section 3  we discuss the existing proposals to solve paradoxes by gravitational effects. Section 4 is devoted to some remarks about the unsolved problems related to Quantum Mechanics, while in the 5th Section we make some general comments on physical theories.

\section{Evolution of a Wave-Packet of the c.m.}

Let us consider as an example the solution of the free 1D Schrödinger equation for an object of mass $M$ in the form
of a Gaussian wave packet (having the ideal uncertainty $\Delta x \Delta p =\frac{1}{2}\hbar $) travelling with a constant average
velocity. It may represent the free motion of the c.m. of a macroscopic object of total mass $M$. Using the Galilean invariance of the free Schr\"odinger equation \cite{Schwinger} one may always go to its own frame (in our
case the c.m. frame of reference) defined by the vanishing of the average velocity and at some initial
time $t = 0$ one has the wave function
{\large
\[
\psi(x,0)=\frac{1}{{(2\pi)}^{\frac{1}{4}}\sqrt{d}}e^{-\frac{x^2}{4d^2}} \qquad .
\]}
of initial width $d$.
Then after a lapse of time $t$ it decays
{\large
\[
\psi(x,t)=\frac{1}{(2\pi)^{\frac{1}{4}}\sqrt{d+\frac{\imath\hbar t}{2dM}}}e^{-\frac{x^2}{4(d^2+\frac{\imath\hbar t}{2M})}} \qquad .
\]
}
This means that the average quadratic width grows as
{\large
\[
\langle x^2 \rangle_{t} =d^2+ (\frac{\hbar t}{2Md})^2 \qquad .
\]
}
 Of course, the
discussion may be extended to the motion in the presence of a slowly varying external potential on the
scale of the macroscopic object as in Ehrenfest’s Theorem. All this is well-known.

One may see, that for $M \to\infty $ the width does not change in time at all. However, the decisive parameter
is not the mass $M$ alone but the product $Md$. For any finite mass $M$ at $d \to 0$ the wave packet decays
instantly. 

Leaving aside the ideal limits, let us consider a numerical example for sake of illustration. Let us consider an $M=1 kg$ object having an initial linear imprecision of its center of mass of $d=0.5 $ \r{A}. This will increase within $10^6$ years indeed only to $0.6$\r{A}.

However, for a small crystal of $M=1 mg$ with the same initial  imprecision of its c.m. this will increase already within 10 years to  $3.28 $\r{A} so that no crystal structure might be "seen" any more. One has to keep in mind, that due to the smeared out c.m. the crystal structure (defined by the quantum mechanical average ion density) the x-Rays encounter is just smeared out and later no structure at all remains. The situation is not equivalent to an oscillating periodic structure.  Had we assumed a much smaller initial width we could get a significant broadening within minutes!  This seems unacceptable.
 
Moreover, this small crystal may have been existed already millions of years ago and still did not spread away.
We must ask
why all macroscopic objects we know have very precise center of mass
position?  

Ultimately, why are there no plane waves of macroscopic objects, although energetically
more favourable. In our example of the Gaussian packet the average kinetic energy is

\[
\frac{\hbar^2}{2M}\langle k^2\rangle=\frac{\hbar^2}{8Md^2} \qquad .
\]
For non-ideal wave-packets $\Delta x \Delta k > \frac{1}{2} $ the average kinetic energy is even greater. Therefore the
more spread is the wave packet, the smaller its conserved average energy is.

No spread out  macroscopic objects
have ever been seen. It looks like a "super-selection" rule for macroscopic objects. "God allowed them
only in this state of precise c.m.?" On the other hand, a world with spread out  mass centres is hard to imagine. It won't function. The macroscopic world, as we know it, is the only possible world!

Further, on a very long time scale no object may be considered as being isolated from its surrounding and the above coherent quantum mechanical description is not any more valid. We are dealing always with open systems in weak contact with a reservoir (thermostat). This one may be ultimately the radiation field.  Therefore the behaviour is turning dissipative over time (see for example \cite{Banyai}) and it may be described only as a mixed ensemble by the quantum-mechanical  density matrix approaching equilibrium while the non-diagonal matrix elements vanish (unless a phase transition occurs!).  This asymptotic equilibrium (predicted by Thermodynamics) in the case of a free particle is a mixed state (canonical distribution) of plane waves extending into the whole space. This is not at all more satisfactory.

 No doubt, the c.m. of a hydrogen atom obeys Quantum Mechanics, but does it obey  also the c.m of this small macroscopic crystal? The answer seems to me  negative. Then how do we define a "macroscopic" object?

On the other hand we are confronted with the extraordinary success of the solid state physics. By the quantum  mechanical treatment of the constituent particles of macroscopic objects we understood the properties of semiconductors, metals and isolators. These touch just the relative motion of the particles. 
It seems that we have to  accept a "double standard" approach. While the relative motion of the ions and electrons in a solid obey quantum mechanics, the center of mass, as well as the overall motion in external fields obey classical physics!

\section{Gravitational proposals.}

An interesting proposal for our sakes is a possible gravitational self-trapping of the c.m. wave packet. A non-linear modification of the Schrödinger equation by gravitation could destroy simultaneously  the superposition principle for macroscopic objects (big masses) without affecting the Quantum Mechanics of microscopic ones. In the same time this might enable to define  the range of validity of pure Quantum Mechanics depending on the mass and dimension of the system. 

Károlyházi \cite{Karolyhazi}\cite{KarolyFrenLuk} (see also \cite{Frenkel} and  Penrose \cite{Penrose}) suggested a modification of Quantum Mechanics due to the gravitational  deformation of the space-time by the constituent particles of the given body in the frame of the general relativity theory. However, the implementation of general relativity into the non-relativistic Quantum Mechanics is highly problematic. 

An introduction of  the Newtonian gravitational attraction between the constituent point-like particles of a system in a local way as by Coulomb interactions, would imply only irrelevant increase, respectively decrease of the Coulomb attraction or repulsion. 

A reasonable alternative would be a gravitational interaction by the mass distributions of the particles defined by the wave function. This implies  non-local, non-linear terms in the Hamiltonian. 

Diósi \cite{Diosi} wrote explicitly the corresponding non-linear many-body Schrödinger equation in the configuration space of N particles, including also self-interaction 
\[
\imath \partial\Psi(X,t)/\partial t =
\left(-\sum_{r=1}^N  \frac{\hbar^2}{2m_r} \frac{´\partial^2}{\partial x_r^2 }                      
-G\sum_{r,s=1}^N\int \frac{m_r m_s}{|x'_s -x_r|}|\Psi(X',t)|^2 d^{3N}X')\right)\Psi(X,t)
\]
Here $X$ denotes the assembly of all particle coordinates $x_r$ of masses $m_r$ ($r=1..N$) and $G$ is the gravitational constant.
 For sake of simplicity we omitted here the Coulomb terms. 

Unfortunately, due to the self-interacting terms the c.m. motion is not separable.  Without these terms however the c.m. equation is not affected at all. It is still a free motion.

For a single particle this equation looks as

\[
\hspace{1cm}
\imath\hbar\frac{\partial}{\partial t}\psi(t,\vec{x})=-\frac{\hbar^{2}}{2M}\Delta 
\psi(t,\vec{x})
-GM^2\!\!\int\!\!d\vec{y}\frac{|\psi(t,\vec{y})|^{2}}{|\vec{x}-\vec{y}|}
\psi(t,\vec{x})\quad,
\]

This is similar to the well-known  s.c. Hartree one, however with attraction instead of repulsion. Dias and Figuero \cite{Dias} had proven the existence of solutions of this equation vanishing at
 $|\vec{x}| \to \infty$ for $t\to \infty$. Diosi \cite{Diosi} itself argues for the existence of a stable ground state, whereas its stability remains unclear.
 
Unfortunately, only a numerical solution \cite{Harrison, Giulini}  might shed light on the trapping of an arbitrary initial wave-packet. Up to now these are not yet convincing.
It is useful to remind also that the above equation has also a stationary  plane-wave solution
 of wave vector $\vec{k}$ with the energy
 \[
 \frac{\hbar^{2}k^2}{2M}-\frac{GM^2}{V}\int_V \!\!d\vec{y}\frac{1}{|\vec{y}|} \quad .
 \]
where $V$ is the volume. In the limit $V\to \infty$ the gravitational correction disappears.
Even if this equation would lead to self-trapping, it is hard to consider it  as the c.m. motion of a many-body quantum mechanical system.
It would mean again a "double standard" approach  describing the relative motion by the N-body Schrödinger equation, while adopting another non-linear single particle Schrödinger equation for the c.m.  
motion.  

Moreover, since the c.m. is just a geometrical point and not a  particle with mass, there is no justification at all for the identification of its inertial mass with its gravitational mass. Either the constituent particles have gravitational mass, or the c.m.? Or even both?

There is  another peculiar approach to the self-capture of the  c.m. motion mixing classical and quantum mechanical arguments.  
Bykov \cite{Bykov} considers a macroscopic sphere of radius $R$ and homogeneously distributed mass $M$ and interprets the quantum mechanical distribution of its c.m. as being equivalent to a statistical distribution of spheres according to the quantum mechanical density. The resulting gravitational interaction potential $U(\vec{x})$ between these spheres he considers as defining the quantum mechanical  Hamiltonian of the c.m. motion. 

\[
\imath\hbar\frac{\partial}{\partial t}\psi(t,\vec{x})=-\frac{\hbar^{2}}{2M}\Delta 
\psi(t,\vec{x})+ U(\vec{x})\psi(t,\vec{x})
\] 
with
\[
U(\vec{x})=\frac{3M^2}{4R^3}G \int d\vec{y} (\vec{x}-\vec{y})^2 |\psi(\vec{y})|^2 \quad .
\]
(By the way, he considers only penetrating spheres!)
In the case of the s-wave this equation however is just the linear Schrödinger equation of an oscillator potential. It has of course bounded stationary states. However, the oscillator eigenstates are not attractors. Therefore no trapping may occur.

 We may conclude that despite their attractiveness gravitational self-trapping  arguments do not offer yet a solution neither for the absence of the broadening of the c.m. nor for the absence of superposition states of macroscopic objects.

 One may still argue that without any modification of the Newtonian point-like attraction between the constituent particles all macroscopic objects are trapped in external
fields, at least in a gravitational one, and the whole discussion about the free c.m. motion is artificial. In a slowly varying potential over the size of a macroscopic object one may expect that in a good approximation the c.m. motion is just the motion of a point-like particle in that potential. While its motion is bounded, a widening remains still unavoidable. 

\section{Discussion}
There is another striking old problem about the application of quantum-mechanical concepts to macroscopic objects. There are no known superposition of states of macroscopic systems. It was  illustrated by the famous "Schr\"odinger's cat". In this respect it is important to mention that the superposition principle remains valid also for open quantum-mechanical systems if their density matrix  obeys linear quantum-mechanical Master equations (see the old review by Gorini at al. \cite{Gorini} and a recent book \cite{Banyai}  about their derivation and properties) .

Another paradox, never mentioned before, follows if one admits the possibility of a full quantum-mechanical description of a macroscopic body by a many-body quantum mechanical state of particles. Then the S-matrix allows transition from a two-particle state to this one.  The probability of such transition may be expressed then as a weighted  many-body free particle transitions. A consequence of this would be that extremely high energy collisions could produce (like all kind of particles or resonances) even arbitrary macroscopic objects. It would be a very surprising event indeed! This again may be related to the mentioned "double standard" approach.

An essential feature of a quantum object is his wave-like behaviour shown by diffraction.  One has diffraction experiments with buckyballs  even greater than in the original experiment \cite{Arndt}, but we cannot expect diffraction with ping-pong balls.

 In view of the discussion in the previous Section, experiments on the c.m. broadening of ultra small pieces of crystals (even in the Angström range!) are also relevant. On our current way toward micro-miniaturization we might get an answer to the question about the threshold between micro and macro. It is very plausible that the macroscopic
physics does not emerge from the microscopic theory. In a mathematical sense it is surely not the  $\hbar \to 0 $ limit.  (See Ref. \cite{Maslov} for the quasi-classical limit of Quantum Mechanics.)

A more limited interpretation in which macroscopic systems cannot be fully described in terms of quantum mechanics of their constituents, should eliminate also the "bootstrap" aspect of Quantum Mechanics i.e. the existence of macroscopic laws as prerequisite for its formulation and application, an aspect worried many physicists.

In a recent review by Bassi et al. \cite{Bassi} discuss the problems of the standard Quantum Mechanics in its application to macroscopic bodies and the manifold of proposals for their solution.  They insist mostly on the violation of the  superposition principle by macroscopic objects and the understanding of the measurement process. We advice the readers to follow their wide bibliography.

 A relatively popular proposal to reconcile the collapse of the wave function during the measurement with the Schrödinger evolution  is the multiple world solution  of Everett \cite{Everett}. It reminds me the very old paper of London and Bauer \cite{London}. Everett introduces the second world concept instead London's solipsist "introspection of the observer". The multiple world concept is however just a  consistent formulation that does not modify the predictions of ordinary quantum mechanics. It does not solve the problem of the wave packet spreading. 
  
 Classical, macroscopic dynamics does not follow from an underlying quantum-mechanical theory. This opinion I share with Basssi et al. \cite{Bassi} who describe also necessary features of an expected new theory. A reconciliation might be found only in a higher theory hidden behind \cite{Ludwig}.
Another important statement by Ref. \cite{Bassi}  I  share  is that "decoherence"  during the measurement cannot explain the lack of superposition of macroscopic objects.  At least, the  loss of coherence due to the contact with the environment (thermostat) that touches our problem of the wave packet, does not destroys the superposition principle. This  follows from the linearity of  quantum-mechanical Master equations for open systems as I remarked before.

  In my understanding Quantum Mechanics has a limited domain of validity like any other known physical theory. The range of this "quantum domain" has yet to be defined by experiments or new theories.  These have to reconcile the  "double standard" description of macroscopic bodies we mentioned in Section 2.

\section{Final Comments}

I am aware that these concepts are somewhat radical. In our human-centred world-image motivated by the exceptional scientific progresses of the last centuries we cheer to construct a unified picture of the world without loopholes and crevices. One looks even for "God's equations" forgetting that mathematics has to be connected to experiment by some interpretation and this is the most difficult one. Any understanding is based finally on words, but words are based on our common macroscopic experience. 

However, all theories of physics of the past were just fragments of knowledge (models) with loose connections between them.  The success of these theories was due to the possibility of simplifying some systems and phenomena  as isolated ones.  Thermodynamics was built up much before statistical mechanics and their link is still rather
vague. The very successful Boltzmann  equation, as well as hydrodynamics have never got a microscopic proof. In the modern  functional integral formalism of 
quantum field theory of elementary particles there is no place for wave functions, Hamiltonians  and observables (operators) while its mathematical structure is highly questionable. It produces Green Functions or S-Matrix of a non-existing operator Quantum Field Theory. Nevertheless, it succeeded to predict the fauna of elementary particles. On the other side, it gives by no means a foundation for the non-relativistic many-body quantum theory.  Its connection to the ordinary quantum mechanics of   systems with  finite degrees of freedom remains a question of belief (see also \cite{Banyai1} ). Efforts to develop a Quantum Mechanics of systems with an infinity of degrees of freedom (C* algebra \cite{Bru}) got yet no new experimental predictions nor general acceptance. The unification of gravitation and quantum theory looks merely a dream. We have different successful theoretical models for different domains of physics. Between them there is only a loose connection, mostly it is only a hope.

Theoretical physics never had a strict vertical structure. It was never derived upwards from the "underlying" theories of matter.  I am afraid, that we are often misled by our need for harmony and exaggerated confidence in our brains.

\vspace{6pt}

{\bf Acknowledgments} The author thanks Mircea Bundaru$\dagger$, Paul Gartner and Sorin Marculescu for fruitful discussions on this topic and for reading the manuscript. Thanks also to an old friend (H.B.) for his wise, sceptical remark: "We understand nothing, nevertheless we explain everything."


\end{document}